\documentstyle[multicol,aps,epsfig,prb]{revtex}

\begin{document}
\twocolumn[\hsize\textwidth\columnwidth\hsize\csname
@twocolumnfalse\endcsname

\date{\today}
\title{Critical properties and phase diagram \\
of the quantum anisotropic $XY$ spin chain in a random magnetic field:\\ 
A density-matrix renormalization-group analysis}

\author{A. Juozapavi\v{c}ius,$^1$ L. Urba,$^1$
S. Caprara,$^{1,2,3}$ and A. Rosengren$^1$}

\address{
$^1$Condensed Matter Theory, Department of Physics, Royal Institute of
Technology, SE--100 44 Stockholm, Sweden\\
$^2$Dipartimento di Fisica, Universit\`a di Roma ``La Sapienza'', P.~le A. 
Moro 2, I--00185 Roma, Italy\\
$^3$Istituto Nazionale di Fisica della Materia, Unit\`a di Roma 1, P.~le A. 
Moro 2, I--00185 Roma, Italy
}

\maketitle

\begin{abstract}
The spin-1/2 quantum anisotropic $XY$ spin chain in a transverse random
magnetic field parallel to the $z$ axis is numerically studied by
means of the density-matrix renormalization group. The dependence of
the spontaneous magnetization and the energy gap on both the strength
of the random magnetic field and the factor of anisotropy is
determined. The critical line for the order-disorder phase transition
is obtained and the resulting phase diagram is drawn. Our results are
compatible with the fact that models with different factors of
anisotropy fall within the same universality class as the quantum
Ising model in a transverse random field.
\end{abstract}
\vskip 0.3truecm]

\section{Introduction}

The effect of disorder on quantum systems is more intriguing than its
classical counterpart, especially when the random noise does not
directly affect the interactions in the original system, but is
coupled to other degrees of freedom, which have nontrivial
commutation relations with the degrees of freedom appearing in the
original Hamiltonian. A simple example within this class of systems is
provided by the transverse randomness in an easy-axis or easy-plane
quantum spin chain. Indeed, when the exchange interactions involve the
$x$ and $y$ spin components, a random magnetic field coupled to the
$z$ spin component produces dramatic effects due to the nontriviality
of the angular-momentum algebra.

In this paper we analyze the one-dimensional spin-1/2 quantum
anisotropic $XY$ model in a transverse random magnetic field, which is
defined by the Hamiltonian
\begin{eqnarray}
{\cal H}=&-&J\sum_{\ell=1}^{N-1} S_{\ell}^x S_{\ell+1}^x-
\gamma J \sum_{\ell=1}^{N-1} S_{\ell}^y S_{\ell+1}^y
\nonumber\\
&-&\sum_{\ell=1}^N h_{\ell} S_{\ell}^z - 
h_x \sum_{\ell=1}^N ({\rm sgn}J)^{\ell} S_{\ell}^x,
\label{hamilton}
\end{eqnarray}
where $N$ is the number of sites in the chain, 
$$
\matrix{ 
S_{\ell}^{\alpha}=&
  \sigma^0 &\otimes&\ldots&\otimes&\sigma^{\alpha}/2&\otimes&\ldots&
\otimes&\sigma^0\cr
~&    1    &  ~    &\ldots&   ~   &   \ell          &   ~   &\ldots&
 ~     &   N
}
$$
are the spin-1/2 operators acting on the Hilbert space spanned by the
vectors $|\psi_1,\ldots,\psi_{\ell},\ldots,\psi_N\rangle$, $\{
\psi_{\ell}\}$ are two-component spinors, $\sigma^0$ is the $2\times
2$ identity matrix, $\{\sigma^{\alpha} \}$ are the $2\times 2$ Pauli
matrices, $\otimes$ is the symbol for tensor product, $J$ is the
coupling constant for the $x$ spin components, $\gamma$ is the factor
of anisotropy, $\gamma J$ is the coupling constant for the $y$ spin
components, and $\{ h_{\ell}\}$ are the on-site values of a random
magnetic field directed along the $z$ axis, which are uniformly
distributed in the interval $-h_0\le h_{\ell}\le h_0$. In the
following we call $h_0$ the strength of the random magnetic field.

Due to the conventions adopted in Eq.~(\ref{hamilton}) a positive
(negative) coupling constant stands for a ferromagnetic
(antiferromagnetic) coupling.  Correspondingly the external field
$h_x$ directed along the $x$ axis, which we include in our model to
force spontaneous symmetry breaking along the easy axis of
magnetization, is uniform (staggered) when $J$ is ferromagnetic
(antiferromagnetic).

In principle the coupling constant $J$ and the factor of anisotropy
$\gamma$ have arbitrary signs, and the only implicit limitation in
Eq.~(\ref{hamilton}) is $-1\le \gamma\le 1$, which amounts to defining
the $x$ and $y$ axes according to the relative strength of the
exchange interaction. However the local gauge transformations
$S_{\ell}^x \to ({\rm sgn}J)^{\ell} S_{\ell}^x$, $S_{\ell}^y \to ({\rm
sgn}J{\rm sgn}\gamma)^{\ell} S_{\ell}^y$, and $S_{\ell}^z \to ({\rm
sgn}\gamma)^{\ell} S_{\ell}^z$ make all the coupling constants
ferromagnetic and the external field $h_x$ uniform, while preserving
the commutation relations among the spin operators, i.e. the
orientation of the local reference frames is preserved by the
transformations. We take then the units of energy such that $J=1$, and
only consider the case $0 \le \gamma \le 1$. We also assume that
$h_x\ge 0$. It is important to observe that the gauge transformations
change the random field as $h_{\ell}\to ({\rm sgn} \gamma)^{\ell}
h_{\ell}$, which is again uniformly distributed in the interval
$[-h_0,h_0]$. However, unlike the Ising limit ($\gamma=0$), at finite
$\gamma$ there is no freedom left to make all the $\{ h_{\ell}\}$
positive (cf. Ref.~\onlinecite{JCR}), because any further local gauge
transformation will not keep fixed both the sign of the coupling
constants and the orientation of the local reference frames.

The outline of the paper is the following. In Sec. II we briefly
recapitulate some previously established results in the two limiting
cases: the $XY$ model in the absence of randomness and the quantum Ising
model in a transverse random field. The main scope of Sec. II is to
provide a framework to which to refer in the discussion and the
interpretation of the results obtained in this paper. In Sec. III we
discuss some technical aspects which are fundamental to obtain
well-defined numerical results, and their relation to the physical
properties of the model. This discussion is essential to allow for the
reproducibility of our results, and to shed light on the physics
underlying the DMRG algorithm. In Sec. IV we determine the critical
properties and the phase diagram of the model (\ref{hamilton}) in the
$h_0$ vs $\gamma$ plane.  A summary of the principal results and some
concluding remarks are found in Sec. V.

\section{Some limiting cases}

The model (\ref{hamilton}) with a generic $\gamma$ is integrable in
the absence of randomness and external magnetic field ($h_0=0$,
$h_x=0$)\cite{lieb} by fermionization of the spin
operators.\cite{noint} The particle-hole excitation spectrum of the
corresponding fermionic system shows, in particular, that the ground
state is separated from the first excited state by an energy
gap\cite{diff}
\begin{equation}
G_0 = 1 - \gamma
\label{five}
\end{equation}
(in units of $J$). In the thermodynamic limit the spectrum above the
gap is continuous and the ground-state energy per site is
\begin{equation}
{\cal E}_0^{{\rm theo}} = - (1+\gamma) \int_{-\pi}^{\pi} {dk\over
8\pi} \left[ 1 - \frac{4\gamma\sin^2 k}{(1+\gamma)^2} \right]^{1/2}.
\label{six}
\end{equation}
A pictorial representation of the physics of the $XY$ model is provided
by the spin wave (SW) theory, which is qualitatively correct (in a
sense that will be clearer in the following), except for $\gamma$ very
close to 1. Expressing the spin operators in terms of auxiliary boson
operators $a_{\ell},a_{\ell}^+$ according to $S_{\ell}^x={1\over
2}-a_{\ell}^+ a_{\ell}$, $S_{\ell}^y={1\over 2}(a_{\ell}^+ +
a_{\ell})$, and retaining only bilinear terms, the Hamiltonian
(\ref{hamilton}) is cast in a diagonal form by a Fourier transform
from real-space to $k$-space representation, and by a subsequent
canonical Bogoliubov transformation. The resulting SW excitation
spectrum is $\varepsilon_k=\sqrt{1-\gamma\cos k}$, and the
ground-state energy per site in the thermodynamic limit is
$$
{\cal E}_0^{\rm SW}=-{3\over 4}+\int_{-\pi}^\pi {dk\over 4\pi} \sqrt{1-
\gamma\cos k}.
$$
I.e., the gap in the excitation spectrum is $G_{\rm SW}=\sqrt{1-\gamma}$,
which coincides with (\ref{five}) only in the Ising limit $\gamma=0$
and in the isotropic limit $\gamma=1$, and is not even perturbatively
correct at small $\gamma$. Nonetheless the discrepancies are mostly
quantitative and the SW theory may be adopted as a reference theory to
discuss the physics of the $XY$ model. In particular, at $\gamma>0$, the
SW Hamiltonian includes a contribution from the zero-point motion of
elementary excitations. The reduction of the spontaneous magnetization
with respect to the saturation value $M_{0,max}={1\over 2}$ in the
Ising limit ($\gamma=0$) is understood within SW theory as the effect
of such quantum fluctuations for $\gamma>0$. Indeed the SW result for
the spontaneous magnetization is
\begin{equation}
M_0={1\over 2}-\int_{-\pi}^\pi {dk\over 8\pi} \displaystyle{
2-\gamma\cos k -2\sqrt{1-\gamma\cos k}\over \sqrt{1-\gamma\cos k} }.
\label{m0sw}
\end{equation}
This result is meaningless at $\gamma=1$ due to the divergence of the
integral in the r.h.s., a fact which is used as an evidence against
the existence of a spontaneous magnetization in the isotropic $XY$
model. However, due to the logarithmic nature of the divergence, it
turns out that the reduction of the magnetization associated with SW
zero-point motion is an underestimate of the reduction in the exact
ground state (see Eq.~(\ref{mg}) and text below it), except for
$\gamma$ very close to 1. Thus in the following, for the sake of
definiteness, we adopt the SW language and we conventionally refer to
the effects of the anisotropy as resulting from quantum fluctuations
with respect to an Ising-like ground state, as it would be in a more
refined (i.e. beyond SW) field theoretical approach to the $XY$ model.

As a consequence of the gap in the excitation spectrum the quantum
fluctuations are massive. The spins are ordered and the system is
ferromagnetic (FM) for $\gamma<1$. In the isotropic limit $\gamma=1$
(the so-called $XX$ model), the gap closes and the spontaneous
magnetization is driven to zero, by massless phase fluctuations. The
system is thus paramagnetic (PM). It is worth mentioning that in the
isotropic limit the model is equivalent to a model of hard-core
bosons, and the absence of spontaneous magnetization in the magnetic
system is related to the absence of Bose-Einstein condensation in the
corresponding one-dimensional interacting boson system.

The issue addressed in this paper is the evolution of the magnetic
ground state for $\gamma<1$ in the presence of the transverse
randomness introduced by a random magnetic field along the $z$
axis. In particular we want to study the interference between the
critical behavior driven by $\gamma\to 1$ and the critical behavior
controlled by the random field $h_0$. Indeed, in the extreme
anisotropic case $\gamma=0$\cite{MW,SM} the system undergoes a FM-PM
quantum phase transition.\cite{fisher} The magnetization at the
critical point $h_0=h_c\equiv {\rm e}/2=1.3529\ldots$ is a
nonanalytical function of the external field $h_x$,\cite{diff1}
$M\left(h_0 =h_c,h_x\right) \sim \left[ \ln \left( D_h/h_x\right)
\right]^{\phi-2}$, where $\phi= (1+\sqrt{5})/2=1.618\ldots$~ is the
golden mean and $D_h$ is a scale factor. The spontaneous magnetization
$M_0\equiv M(h_x\to 0)$ in the ordered phase decreases with increasing
$h_0$, and
\begin{equation}
\label{three}
M_0 \sim (h_c-h_0)^{2-\phi}
\end{equation}
for $h_0\to h_c^-$. Contrary to ordinary continuous phase
transitions, where the nonanalyticity is limited to the critical
point, a Griffiths' region\cite{griffiths} exists around the critical
point, where $M(h_x)$ is not analytical. In the weakly ordered
Griffiths' region ($h_0<h_c$), for small $h_x$,
\begin{equation}
M(h_x)-M_0 \sim h_x^{\alpha}
\left[ \ln\left( D_h'/h_x\right) \right]^\kappa,
\label{four}
\end{equation}
where $M_0$ is the spontaneous magnetization, the exponent $\kappa$
cannot be determined within renormalization group, $D_h'$ is a
nonuniversal scale factor, and the exponent $\alpha$ depends
continuously on the distance from criticality $\delta\sim h_c-h_0$.
\cite{fisher} As far as $\alpha\le 1$, the spin susceptibility
diverges as $h_x\to 0$, i.e. the magnetization curves intercept the
axis $h_x=0$ with an infinite slope, even when $M_0$ is finite.

Previous results of density-matrix renormalization-group (DMRG)
calculations at $\gamma=0$\cite{JCR} were in good agreement both with
the theory\cite{MW,SM,fisher} and other numerical
calculations.\cite{young} Below we extend these results to the entire
region $h_0>0$, and $0\le \gamma\le 1$, and determine the full phase
diagram and the critical properties of the model (\ref{hamilton}). We
are interested in understanding the role of quantum fluctuations (in the
sense clarified above) in the Griffiths' region. The issue is
particularly intriguing for $\gamma\to 1$, when the critical behavior
of the nonrandom system interferes with the critical behavior
controlled by the random field. This interference is directly evident
in those quantities which vanish at criticality, such as the
spontaneous magnetization and the gap in the excitation spectrum.

We point out that, due to the iterative nature of the infinite-size
DMRG algorithm, the system size is increased while the irrelevant
states are progressively truncated away. The discrepancy between the
infinite-size limit of the algorithm and the thermodynamic limit of
the physical system depends on the size of the error introduced by the
truncation.\cite{white} Thus the technical aspects related to the
definition of a proper numerical procedure are deeply entangled with
the physical properties of the system. In particular it will be clear
that, in order to drive the system towards the phase transition, it is
not only necessary to reach the infinite-size limit, in the presence
of an infinitesimal external field, but it is also important to
control and determine the effect of the truncation of the Hilbert
space. The full procedure explained below is essential to reproduce
our results and obtain the correct limiting behaviors.

\section{Algorithm and technicalities}

We perform our calculations using the infinite-size DMRG algorithm
introduced by White\cite{white} with some minor modifications
described in Ref.~\onlinecite{JCR}. Accordingly, we adopted open
boundary conditions, which produce more accurate results within
DMRG.\cite{white}

\begin{figure}[h]
\epsfxsize=7cm
\centerline{\epsfbox{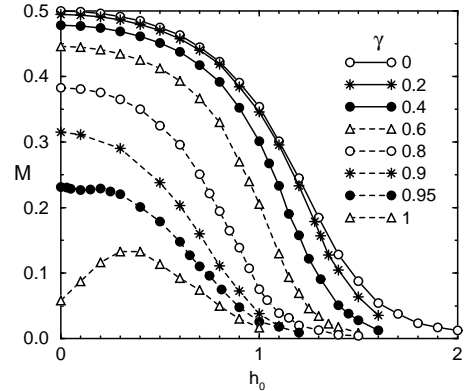}}
\caption{ The dependence of the magnetization on the random magnetic
field for different values of the factor of anisotropy $\gamma$. The
system size is 302 sites, the uniform magnetic field $h_x=10^{-5}$,
and eight states are kept at each DMRG step. The large value of the
magnetization in the disordered region is essentially due to the
anomalous response to the external field $h_x$ in the Griffiths'
region.}
\label{fig1}
\end{figure}

At a fixed number $s$ of states kept in the truncation of the density
matrix and after a (sufficiently large) fixed number $N_{\rm RG}$ of DMRG
steps, when the chain consists of $N=2N_{\rm RG}+2$ sites, the
magnetization at a given (small) uniform magnetic field $h_x$
decreases both with the strength of the random field $h_0$ and the
factor of anisotropy $\gamma$, except for a region close to $\gamma=1$
where the magnetization appears to be a nonmonotonic function of
$h_0$ (see Fig.~\ref{fig1}). The strong dependence of the
magnetization on the external field around the critical point is
essentially the result of a divergent susceptibility in the Griffiths'
region. Even for a field as small as $h_x=10^{-5}$, the transition
point is masked by the strongly nonanalytical behavior, and is only
hinted by the change in the curvature of the $M$ vs $h_0$ curves.

Moreover, it is evident that particular care has to be exercised close
to the isotropic limit $\gamma=1$, to extract the correct
behavior. Indeed, quantum fluctuations suppress the spontaneous
magnetization at $\gamma=1$ even in the absence of randomness. Thus we
expect $M_0$ to be zero for all values of $h_0$. This indicates that a
proper procedure has to be defined, which is reliable and unique for
all $\gamma$. We carefully address this issue and the results are
given below.

To determine the spontaneous magnetization a (small) uniform magnetic
field $h_x$ in the $x$ direction is applied to lift the degeneracy in
the ground state, which is doubly degenerate at $h_x=0$ (this
degeneracy occurs for large system sizes for $\gamma>0$ and for any
size at $\gamma=0$). Usually, for small values of $\gamma$, any $h_x$
is enough to truncate the undesired degenerate state after the initial
DMRG steps. However, the closer one gets to $\gamma=1$, the larger
this uniform field should be, and more than 150 DMRG steps could be
needed to isolate the state with broken symmetry. Further, increased
accuracy of calculations (i.e. a larger number of states kept) is also
needed close to $\gamma=1$. We have checked that the difference
between the exact ground-state energy ${\cal E}_0^{{\rm theo}}$, given
by Eq.~(\ref{six}), and the calculated one ${\cal E}_0^{{\rm DMRG}}$
for a fixed number of states kept in the truncation procedure, $s=8$,
closely follows the truncation error\cite{white}
$1-P_s=1-\sum_{m=1}^{s} \varrho_m= \sum_{m>s}\varrho_m$, i.e. the sum
of the eigenvalues of the density matrix which correspond to the
truncated states. This error increases with increasing $\gamma$, from
$10^{-13}$ at $\gamma=0$, to $10^{-4}$ at $\gamma=1$. This phenomenon
is deeply related to the increasing complexity of the ground state, as
the role of quantum fluctuations is emphasized. As a consequence, the
spontaneous magnetization has a dependence on the parameters $s$ and
$N_{\rm RG}$ which becomes more and more important as $\gamma\to 1$.

To deal with this dependence and obtain the correct spontaneous
magnetization $M_0=M(h_x\to 0)$ for all $\gamma$, the following steps
are needed: (i) We calculate the magnetization along the easy axis
$$
M_j^{(s)}(N_{\rm RG},h_x,h_0)= (2N_{\rm RG}+2)^{-1} \times
$$
\begin{equation}
\langle h_x,h_0;N_{\rm RG},s,j| {\tilde S}^x |h_x,h_0;N_{\rm RG},s,j\rangle,
\label{qav}
\end{equation}
where $\langle\,\cdot\,\rangle$ is the quantum expectation value in
the ground state and ${\tilde S}^x$ is the properly truncated total
$x$-spin operator, at each DMRG step $N_{\rm RG}$, for a given uniform
magnetic field $h_x$ and a random magnetic field of strength $h_0$,
while $s$ states are kept during the DMRG procedure. The index $j$
labels a particular realization of the random field. Then (ii) we
average this value over $N_c$ realizations of the random field,
$\overline{M}^{(s)}(N_{\rm RG},h_x,h_0)=N_c^{-1}\sum_{j=1}^{N_c}
M_j^{(s)}(N_{\rm RG},h_x,h_0)$, where a suitable value for $N_c$ depends
on both $\gamma$ and $h_0$. In most cases an average over $N_c=500$
configurations gives satisfactory results, though sometimes more than
2000 configurations are needed (for $\gamma\simeq 1$ and close to
criticality).

Due to the effects of the edges of the chain, $\overline{M}$ always
increases with increasing lattice size, but the increase is very slow
above some typical chain size ($N_{\rm RG}=50$), and may be fitted as
$\overline{M}^{(s)}(N_{\rm RG},h_x,h_0)=
\overline{M}^{(s)}(\infty,h_x,h_0)-A/ N_{\rm RG}$, where $A$ and
$\overline{M}^{(s)}(\infty,h_x,h_0)$ are fitting parameters, and so
(iii) we extrapolate $\overline{M}$ to the infinite-size limit
$N_{\rm RG}\to\infty$ and obtain $\overline{M}^{(s)}(\infty,h_x, h_0)$.

\begin{figure}[h]
\epsfxsize=7cm
\centerline{\epsfbox{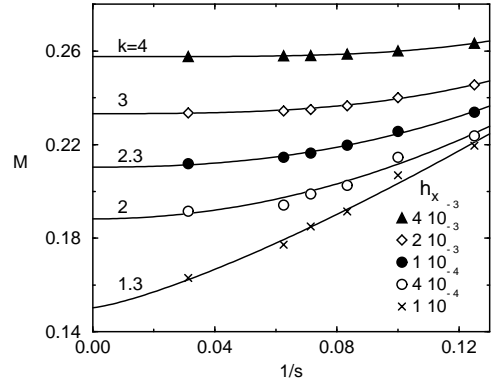}}
\caption{ The dependence of the magnetization (extrapolated to
$N_{\rm RG}\to\infty$) on the inverse number of states kept, $1/s$, at
$\gamma=1$, in the absence of randomness, $h_0=0$. The solid lines are
fits of the form $M=M(0)+B\times (1/s)^k$, the values of the exponent
$k$ are shown.}
\label{fig2}
\end{figure}

The larger the number of states $s$ kept is, during renormalization,
the better the accuracy obtained is, so the magnetization slightly
decreases with $s$ (a qualitative argument is: the smaller the number
of states kept is, the more influential the state with all spins up
is, the larger $M$ is). Thus (iv) we make an extrapolation to
$s=\infty$, \cite{white} using the formula
\begin{equation}
\overline{M}^{(s)}(\infty,h_x,h_0)=
\overline{M}^{(\infty)}(\infty,h_x,h_0)+B\times\left( {1\over
s}\right)^k,
\label{onepower}
\end{equation}
which fits the data quite well (except for small oscillations at
larger $1/s$), and we obtain
$\overline{M}^{(\infty)}(\infty,h_x,h_0)$, (see Fig.~\ref{fig2}). A
more accurate formula is slightly different. On physical grounds,
indeed, one expects that the magnetization
$M(1/s)\equiv\overline{M}^{(s)}(\infty,h_x,h_0)$ consists of two
parts: $M(1/s)=M_{{\rm reg}}(1/s)+M_{{\rm noise}}$, where the regular
part $M_{{\rm reg}}(1/s)$ has a Taylor expansion close to $1/s=0$ and
the noise part $M_{\rm noise}$ is due to nonuniform changes in the
ground-state properties and only exists at large $1/s$. As the field
$h_x$ is increased, the dependence on $1/s$ becomes much weaker (in
strong fields the ground state is fully magnetized and so $s=1$ gives
the exact solution), which means that for large fields one expects
$M(1/s)$ to be a noiseless horizontal line, which can be imagined as
resulting from a successive vanishing of the terms in the Taylor
expansion of $M_{\rm reg}(1/s)$. Indeed, a Taylor expansion up to
fifth order shows that only the fourth order $1/s$ term has any
influence for the $h_x=4\times 10^{-3}$ data (Fig.~\ref{fig2}), though
both the linear and square $1/s$ terms are present for
$h_x=1\times 10^{-4}$. The fact that no more than two nonconstant terms
are ever present in the Taylor expansion for any of the chosen $h_x$
allows us to introduce with sufficient accuracy the effective formula
(\ref{onepower}), which is desirable because of a smaller number of
free parameters. The presence of two (nonconstant) terms in the
Taylor expansion at small $1/s$ results in a noninteger exponent $k$
in Eq.~(\ref{onepower}), where only one power is considered. The
exponent $k$ is intermediate between the two corresponding integer
exponents appearing in the Taylor expansion.

\begin{figure}[h]
\epsfxsize=7cm
\centerline{\epsfbox{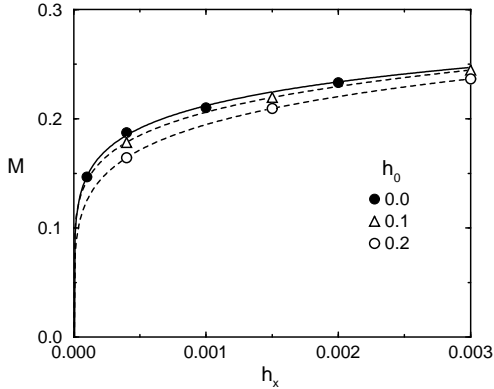}}
\caption{ The dependence of the magnetization (extrapolated to
$N_{\rm RG}\to\infty$, $s\to\infty$) on the uniform magnetic field $h_x$
at $\gamma=1$. The lines are fits of the form $M=M_0+C h_x^{\alpha}$.}
\label{fig3}
\end{figure}

Finally, (v) we extrapolate $\overline{M}^{(\infty)}(\infty,h_x,h_0)$
to zero external field $h_x$, using the expression
$\overline{M}^{(\infty)}(\infty,h_x,h_0)=
\overline{M}^{(\infty)}(\infty,0,h_0)+C h_x^{\alpha}$, where $C$ is a
fitting parameter, $M_0(h_0) \equiv
\overline{M}^{(\infty)}(\infty,0,h_0)$ is the spontaneous
magnetization as a function of $h_0$ (at fixed $\gamma$) and the
exponent $\alpha$ is determined by the fit to numerical data and is
smaller than 1 in the Griffiths' region, where the susceptibility
diverges as $h_x\to 0$. The above formula is a slight modification of
Eq.~(\ref{four}), based on the observation that the exponent $\kappa$
is negligible. The modified formula reproduces the behavior of
numerical data for any value of the factor of anisotropy $\gamma$. As
Fig.~\ref{fig3} shows, the magnetization is decreasing with increasing
$h_0$, and we clearly see that $M_0$ goes to zero at $\gamma=1$ even
though the uncertainty is rather large which is due to that we are
exactly at criticality, e.g. $M_0(h_0=0)=-0.18 \pm 0.18$, when
properly extrapolated.

We point out that the above successive extrapolations are not needed
in the Ising limit ($\gamma=0$) Ref. \onlinecite{JCR} and add but
little accuracy for small $\gamma$. This is due to the fact that the
truncation error decreases very rapidly as a function of the number
$s$ of states kept when $\gamma$ is small. Thus, provided the system
size is reasonably large and $s$ not too small, the only extrapolating
procedure needed to get the correct behavior for the spontaneous
magnetization, close to the Ising limit, is $h_x\to 0$.\cite{JCR}

\section{Results and discussion}

\begin{figure}[h]
\epsfxsize=7cm
\centerline{\epsfbox{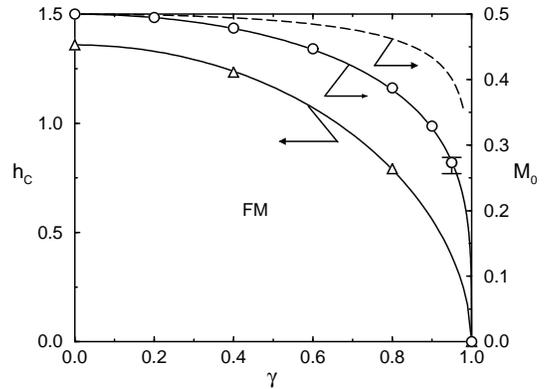}}
\caption{ The triangles represent the critical points $h_c$ obtained
by DMRG calculations for different values of the factor of anisotropy
$\gamma$. The solid line is the best fit of the form
$h_c=h(0)(1-\gamma^2)^\varrho$ with $\varrho=0.52 \pm 0.01$. The
system is in a ferromagnetic state below this line. The circles
represent the dependence of the spontaneous magnetization on $\gamma$
at $h_0=0$. Data are fitted by $M_0 = 0.5 (1-\gamma^2)^\vartheta$ with
$\vartheta=0.255^{+0.010}_{-0.005}$. The error bars that are smaller
than the symbol size are not shown. The result of SW theory,
Eq.~(\ref{m0sw}), is also drawn (dashed line).}
\label{fig4}
\end{figure}

As a starting test for the procedure discussed in Sec. III, we have
calculated the spontaneous magnetization $M_0$ for many different
values of $\gamma$ at zero random field. As we anticipated, quantum
fluctuations tend to reduce the magnetization. We fitted our data as
\begin{equation}
M_0(h_0=0,\gamma) = {1\over 2} (1-\gamma^2)^\vartheta,
\label{mg}
\end{equation}
with $\vartheta=0.255^{+0.010}_{-0.005}$, very close to $1/4$ (the
circles in Fig.~\ref{fig4}), which is to be compared to the SW
result Eq.~(\ref{m0sw}) (the dashed line in Fig.~\ref{fig4}). In SW
language, the factor $(1-\gamma^2)^\vartheta$ represents the reduction
of the magnetization, with respect to the saturation value
$M_0(h_0=0,\gamma=0)=1/2$, due to quantum fluctuations. This factor is
independent of the sign of $\gamma$, and drives the spontaneous
magnetization to zero as $\gamma\to 1$. Thus, even in the absence of
randomness, a critical behavior is induced by quantum fluctuations.
This behavior interferes with the critical behavior controlled by the
random field.

\begin{figure}[h]
\epsfxsize=7cm
\centerline{\epsfbox{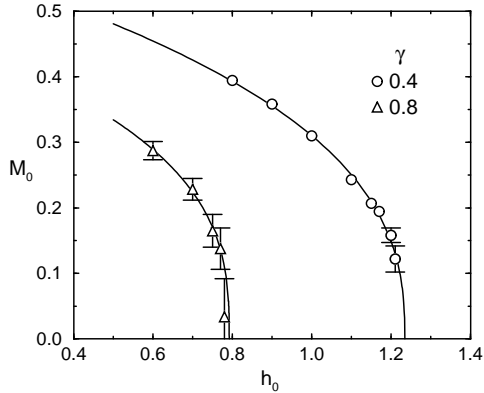}}
\caption{ The dependence of the spontaneous magnetization on the
strength of the random magnetic field $h_0$. The data points were
fitted by the formula $M_0=D\times(h_c-h_0)^\beta$.}
\label{fig5}
\end{figure}

To analyze this interplay, we applied the same procedure to obtain the
spontaneous magnetization in the presence of randomness
(Fig.~\ref{fig5}). Only $s=8$ states proved to be enough to obtain
accurate spontaneous-magnetization data for $\gamma=0.4$ and no
additional extrapolation to $s\to\infty$ was necessary. The above
mentioned procedure was fundamental, instead, to produce sensible data
for $\gamma=0.8$.  About 1000 runs were needed for a given strength of
the random field $h_0$, the uniform field $h_x$ and the number of
states kept $s$ to get an accurate statistical average of the
magnetization. The available computational resources limited us to
$s\le 12$, so it was very important to have the fit formula
(\ref{onepower}) with a small number of free parameters. The accuracy
of the final $M_0$ data is good despite the small number of states $s$
used in the calculations, which induces errors in the exponent $k$ and
consequently in $M(1/s=0)$, (see Eq.~(\ref{onepower}) and the text
below it).

Once the spontaneous magnetization is obtained, the next step is to
determine the phase-transition line $h_c(\gamma)$, i.e. the line in
the $h_0$ vs $\gamma$ plane where the spontaneous magnetization $M_0$
vanishes. Since the data close to criticality are usually affected by
large errors, the best way to obtain $h_c$ is to find $M_0$ in the
ordered phase and then fit the data to formula (\ref{three}), leaving
$\beta=2-\phi$ as an adjustable parameter:
$M_0=D\times(h_c-h_0)^\beta$, as it is shown in Fig.~(\ref{fig5}). The
results are $h_c=1.234 \pm 0.007$ and $\beta=0.38 \pm 0.03$ for
$\gamma=0.4$ and $h_c=0.79 \pm 0.01$ and $\beta=0.35 \pm 0.04$ for
$\gamma=0.8$.  We emphasize here the fact that the magnetization fit
gives an exponent $\beta$ compatible with $2-\phi\approx 0.38$ for any
$\gamma$, which is a strong numerical evidence that the system falls
in the same universality class as the quantum Ising spin chain for all
$\gamma<1$. In the language of SW theory the presence of a gap in the
excitation spectrum leads to an effective purely Ising-like model,
with a renormalized coupling constant $J_\gamma$, near criticality.

The phase diagram is plotted in Fig.~(\ref{fig4}). The critical line
(the one passing through the triangle symbols which correspond to our
numerical results) is well approximated by the equation
\begin{equation}
h_c(\gamma) = h_c(0) (1-\gamma^2)^\varrho
\label{crl}
\end{equation}
where $h_c(0)={\rm e}/2$ and $\varrho=0.52 \pm 0.01$, very close to
$1/2$.  Thus the effect of quantum fluctuations in the Griffiths'
region around the critical point is such as to increase the effective
strength of the random field. In other words, as the magnetic order is
weakened by quantum fluctuations, a weaker random field $h_0$ is
needed to drive the system to the paramagnetic phase.  We point out
that the critical field strength vanishes for $\gamma\to 1$, where a
different critical behavior sets in [cf. Eq.~(\ref{mg})], controlled
by the corresponding critical point.

In the language of SW theory quantum fluctuations lead to a
renormalized effective Hamiltonian. The result (\ref{mg}) indicates
that the field $S_\ell^x$ is renormalized as ${\tilde
S}_\ell^x=(1-\gamma^2)^\vartheta S_\ell^x$ at $h_0=0$. This in turn
leads to an effective Ising-like model with renormalized coupling
constant $J_\gamma=J (1-\gamma^2)^{2\vartheta}$. Thus the result
(\ref{crl}) indicates that the increase in the strength of random
field due to quantum fluctuations is entirely due to a reduction of
the coupling constant, while the coupling to the random field is not
renormalized, so that the scaling law $\varrho= 2\vartheta$ holds. In
other words, the interference of the two critical behaviors leads to
the relation $(2M_0)^2=2h_c/{\rm e}$, i.e. the ratio
$M_0^2/h_c=1/2{\rm e}$ is independent of $\gamma$.

\begin{figure}[h]
\epsfxsize=7cm
\centerline{\epsfbox{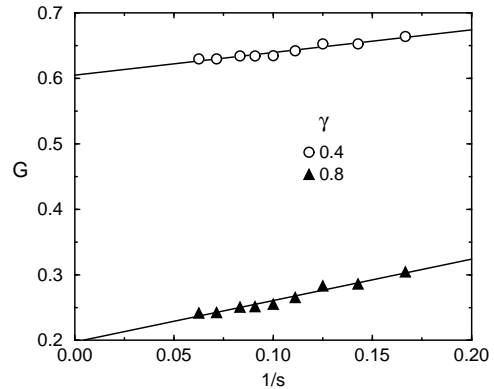}}
\caption{ The dependence of the energy gap on the inverse number of
states kept in the DMRG truncating procedure. The system size is up to 404
sites, $h_x=0$, $h_0=0$. The straight lines are fits of the form
$G=a+b/s$, where $a=0.605 \pm 0.003$ for $\gamma=0.4$ and $a=0.197 \pm
0.004$ for $\gamma=0.8$.}
\label{fig6}
\end{figure}

To check the consistency of the phase diagram we have calculated the
energy gap $G$ between the ground state and the first excited state,
which provides an independent determination of the critical point.
Indeed, randomness tends to fill in the gap, which should become zero
at criticality. This is a property, which is easier to calculate,
compared to the spontaneous magnetization, because it can be
calculated at zero uniform magnetic field and no extrapolation for
$h_x\to 0$ is needed.  However, unlike the ground-state properties,
the properties of the excited states require the full extrapolating
procedure $N_{\rm RG}\to \infty$, $s\to \infty$, to be properly
determined, even in the Ising limit.  Indeed the energy gap depends
strongly on the number $s$ of states kept in the DMRG procedure for
all $\gamma$. Some examples are shown in Fig.~\ref{fig6}, in the
absence of randomness. It is seen that, once the correct extrapolation
procedure is adopted, the DMRG reproduces the exact result
(\ref{five}) within good accuracy.

\begin{figure}[h]
\epsfxsize=7cm
\centerline{\epsfbox{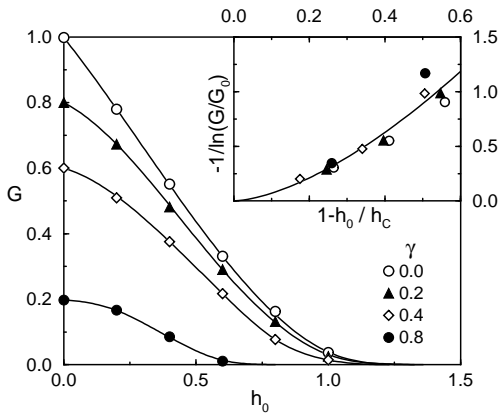}}
\caption{ The dependence of the energy gap on the strength of the
random field $h_0$. The system size is above 200 sites, $h_x=0$. The
average is performed over 500 realizations of the random field. The
inset shows rescaled data, and the solid curve is $-1/\ln
(G/G_0)=K(1-h_0/h_c)^\lambda$ with $K=2.6\pm0.4$ and
$\lambda=1.6\pm0.2$. $G_0$ denotes the gap in the absence of
randomness, Eq.~(\ref{five}).}
\label{fig7}
\end{figure}

We have calculated the energy gap $G$ for different $\gamma$ and $h_0$
(Fig.~\ref{fig7}). As $h_0$ increases, $G$ is reduced from its initial
value (\ref{five}) and is driven to zero as the critical point is
approached. The results for the gap are in agreement with the results
obtained from the spontaneous magnetization data (cf.~the triangles in
the Fig.~\ref{fig4}). The curves are exponentially flat near
criticality so that the critical points are much more accurately
determined from the magnetization curves. However, once $h_c$ is
determined from magnetization data, the data corresponding to
$-1/\ln(G/G_0)$ collapse near criticality on one curve as a function
of $1-h_0/h_c$, as shown in the inset of the Fig.~\ref{fig7}.

So far, we have discussed the procedure to obtain the critical
behavior of the anisotropic $XY$ model in a random field, and have
studied those quantities which vanish at criticality. Now we analyze
the correlation length and we look for the nearly critical behavior at
a (large) fixed system size and a given number of states in the
truncation procedure kept.  In particular, since there is a great deal
of analysis on finite-size effects in approximate numerical
calculations for models near criticality, we concentrate on the
distinguishing feature of the DMRG method, i.e. the effect of the
truncation of the Hilbert space.  Thus we calculated both the
connected and nonconnected two-point spin-spin correlation functions
along the easy axis of magnetization
\begin{eqnarray}
C_j^t(r)&=&4\left[ \langle S_0^x S_r^x\rangle-
\langle S_0^x\rangle\langle S_r^x\rangle \right]
\label{cfcn}\\
C_j(r)&=&4 \langle S_0^x S_r^x\rangle,
\label{cfucn}
\end{eqnarray}
where $\langle\,\cdot\,\rangle$ stands for the quantum expectation
value in the ground state, for any given realization $j$ of the random
field, and we are using here an abbreviated notation with respect to
Eq.~(\ref{qav}).  A factor of 4 in the definition of $C_j(r)$ was
introduced to allow for a direct comparison with
Ref.~\onlinecite{fisher} in the Ising limit $\gamma=0$.  We point out
that, for a given realization of the random field, $\langle
S_0^x\rangle$ and $\langle S_r^x\rangle$ are different, and neither of
them is equal to the magnetization given by Eq.~(\ref{qav}), due to
the lack of translational invariance.

We used a method which is slightly different from (but essentially
equivalent to) the method described, e.g., in
Ref.~\onlinecite{JCR}. We grew the system up to a size of $\sim 400$
sites, while storing the truncated spin operators for the different
sites of the chain. The typical $C_j^t(r)$ should decay exponentially
with a correlation length $\xi$.\cite{fisher} The plot of the average
over disorder $N_c^{-1}\sum_{j=1}^{N_c} \ln C_j^t(r)$ as a function of
the distance $r$ allows then to extract the harmonic
\begin{figure}[h]
\epsfxsize=7cm
\centerline{\epsfbox{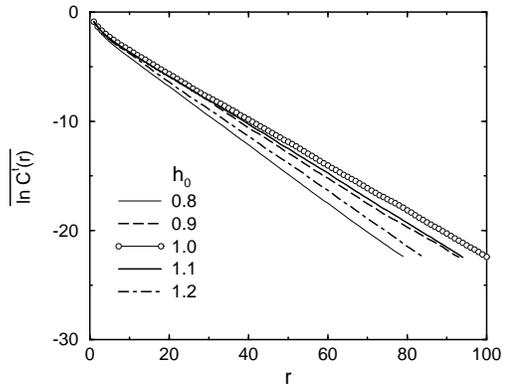}}
\caption{ The average (over 500 configurations) of the logarithm of
the connected correlation function for $\gamma=0.8$. The typical
correlation length $\xi$ is calculated as the negative inverse slope
of these curves.}
\label{fig8}
\end{figure}
\noindent
average of the correlation length from the slope of the linear
behavior at large $r$ (see, e.g., Fig.~\ref{fig8}). By plotting the
$\xi$ vs $h_0$ for each $\gamma$ (Fig.~\ref{fig9}), we obtained the
peaks that indicate the would-be critical point. As it is evident, the
peaks are rather broad, and the correlation lengths rather short, due
to the truncation of the Hilbert space, but the position of the
maximum can be extracted with a good accuracy, and we found $h_{\rm
peak}=1.40\pm 0.05$ for $\gamma=0.0$, $h_{\rm peak}=1.25\pm 0.05$ for
$\gamma=0.4$, and $h_{\rm peak}=1.00\pm 0.05$ for $\gamma=0.8$. We
checked that increasing the number of states $s$ kept in the
truncating procedure leads to narrower peaks which shift towards the
critical values $h_c(\gamma)$ that were obtained from the spontaneous
magnetization data in the infinite-size and $s\to\infty$ limit.

\begin{figure}[h]
\epsfxsize=7cm
\centerline{\epsfbox{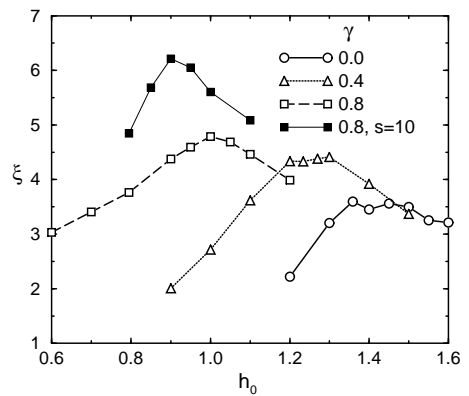}}
\caption{ The dependence of the typical correlation length $\xi$ on the
strength of the random magnetic field $h_0$. There were $s=8$ states
kept in the truncation procedure for the white data points.}
\label{fig9}
\end{figure}

\begin{figure}[h]
\epsfxsize=7cm
\centerline{\epsfbox{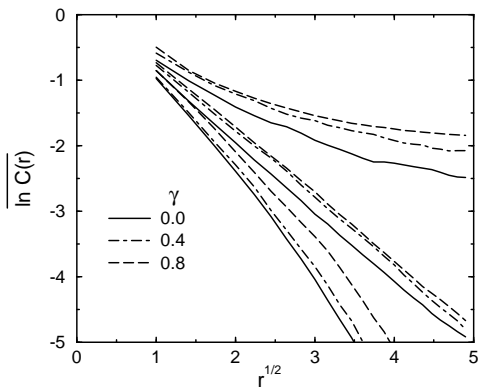}}
\caption{ The average of the logarithm of the nonconnected
correlation function. The curves correspond to the following strengths
of the random magnetic field (from top to bottom): $h_0=1.25;\ 1.44; \
1.65$ for $\gamma=0.0$ (solid lines), $h_0=1.1;\ 1.3;\ 1.5$ for
$\gamma=0.4$ (dot-dashed lines), and $h_0=0.8;\ 1.03;\ 1.2$ for
$\gamma=0.8$ (dashed lines).}
\label{fig10}
\end{figure}

To see how close to criticality the system is at the peaks of the
correlation lengths we looked for a typically critical property of the
correlation functions. In Fig.~\ref{fig10} we plotted the average over
the realizations of the random field of the logarithm of the
correlation function (\ref{cfucn}), $\overline{\ln C(r)} \equiv
N_c^{-1}\sum_{j=1}^{N_c} \ln C_j(r)$ as a function of $\sqrt{r}$,
since such a behavior is expected at criticality.\cite{JCR,fisher} Our
result shows that the curves are very sensitive to the value of $h_0$,
and bend upwards when $h_0$ is smaller than some particular value
$\hat{h}$ or downwards when $h_0>\hat{h}$. The best fit to linear
behavior is found at $\hat{h}=1.44$ for $\gamma=0.0$, $\hat{h}=1.30$
for $\gamma=0.4$, and $\hat{h}=1.03$ for $\gamma=0.8$. These values
and the corresponding $h_{\rm peak}$, obtained from the $\xi$ vs $h_0$
curves, coincide within the error bars. Thus the truncated Hilbert
space embodies the critical behavior of the system in a
self-consistent though approximate way. Increasing the size of the
Hilbert space improves the accuracy in the description of the critical
properties until the only limitation to a fully developed criticality
is the finite size of the system. It is perhaps important to remark
that the extrapolating procedure proposed in Sec. III to extract the
spontaneous magnetization may also be applied to the correlation
functions. The only difference is that much more computational time
and memory is required to calculate and store the data which refer,
for a given distance $r$, to different system sizes, different number
of states $s$, and different random field realizations.

\section{Conclusions}

In summary, we analyzed some properties of the spin-1/2 quantum
anisotropic $XY$ chain in a transverse random magnetic field by means of
the density-matrix renormalization group. The dependence of the
magnetization on the uniform magnetic field $h_x$, the
strength of the random magnetic field $h_0$ and the factor of
anisotropy $\gamma$ was obtained. The order-disorder phase-transition
line was determined [Eq.~(\ref{crl})] and the phase diagram was
drawn. The energy gap between the ground state and the first excited
state was investigated as well. The dependence of the gap on both the
strength of the random field $h_0$ and the factor of anisotropy was
obtained. The gap vanishes at the phase transition, determined
independently from magnetization data, and reproduces the correct
limiting value (\ref{five}) as the strength of the random field is
reduced. Finally we calculated the connected and nonconnected
two-point spin-spin correlation functions along the easy axis of
magnetization. We studied in particular the asymptotic behavior for
large distances at criticality, and found that the $\sqrt{r}$-behavior
found at $\gamma=0$\cite{JCR,fisher} persists for $\gamma>0$.

The critical properties are remarkably the same for all $\gamma<1$,
providing clear numerical evidence for universality, i.e. models with
different values of the factor of anisotropy $0 \le \gamma < 1$ all
belong to the same universality class as the spin-1/2 quantum Ising
chain in a transverse random magnetic field.

The main advantages of the method used in this paper to investigate
the properties of a random quantum system, with respect to other
numerical methods have been discussed in our previous paper.
\cite{JCR} Here we wish to comment in deeper detail on the specific
technical problem which was dealt with in this paper. Indeed, in the
present case, the density-matrix renormalization-group approach
requires a particular care, due to the interplay of randomness,
finite-size effects, and because the consequences of truncation of the
Hilbert space are enhanced by the presence of quantum fluctuations
(with respect to an Ising-like ground state) as $\gamma$ is
increased. This effect is most easily seen in the gradual spread of
the eigenvalues of the density matrix, i.e. in the increasing
importance of including more states in the truncating procedure, in
order to obtain an accurate description of the system. However a
reliable protocol was discussed in this paper, which takes care of all
those aspects on equal footing, allowing us to obtain physical results
which are very robust with respect to ``local'' variations of the
extrapolating procedures.

Criticism was often raised against the use of the DMRG in the
investigations of random systems (see, e.g. Ref.~\onlinecite{kur}),
and some modifications of the DMRG procedure were proposed to deal
with randomness.\cite{hida} The authors essentially refer to the
failure in accommodating sudden changes of the ground state within a
truncated basis for the Hilbert space. These objections are
appropriate, in principle, and suggest a careful analysis of the
stability of the DMRG results as the size of the basis is enlarged. In
this paper we showed that this analysis is possible, and the
objections may be overcome.

\acknowledgments

This work was supported by The Swedish Natural Science Research
Council, and with computing resources by the Swedish Council for
Planning and Coordination of Research (FRN) and Parallelldatorcentrum
(PDC), Royal Institute of Technology, Sweden. One of us (SC)
acknowledges partial financial support of the I.N.F.M. - P.R.A. 1996,
many useful discussions with Dr. S. De Palo and Dr. N. Cancrini, and
an illuminating suggestion from Prof. C. Di Castro.


\begin{references}
\bibitem{noint} 
It is worth mentioning that the $XY$ model is nonintegrable when $h_x$ is 
finite.
\bibitem{diff}
We adopt in our paper a notation which is different from the one
adopted in Ref.~\onlinecite{lieb} in both the definition of the spin
operators, and of the anisotropy factor. All the results have been
consistently translated into our notation.
\bibitem{diff1} The notation here is different from both the notation
adopted in our previous paper\cite{JCR} and the notation used in
Ref.~\onlinecite{fisher}.  This is due to the fact that the model
discussed in the present paper is known in the literature as the $XY$
model, whereas the Ising limit is usually discussed within a $Z$
model. Thus the $x$ and $z$ axes of Refs.~\onlinecite{JCR,fisher} have
been interchanged here.
\end{references}
\end{document}